\documentstyle[prd,aps,preprint,tighten,epsfig]{revtex}

\begin{document}

\draft

\title{Generalized Tri-bimaximal Neutrino Mixing \\
and Its Sensitivity to Radiative Corrections}
\author{{\bf Shu Luo} ~ and ~ {\bf Zhi-zhong Xing}}
\address{CCAST (World Laboratory), P.O. Box 8730, Beijing 100080,
China \\
and Institute of High Energy Physics, Chinese Academy of Sciences,
\\
P.O. Box 918, Beijing 100049, China
\footnote{Mailing address} \\
({\it Electronic address: xingzz@mail.ihep.ac.cn}) }
\maketitle

\begin{abstract}
We argue that the tri-bimaximal neutrino mixing pattern $V_0$ or
its generalized form $V'_0$, which includes two arbitrary Majorana
phases of CP violation, may result from an underlying flavor
symmetry at a superhigh energy scale close to the seesaw scale
($\sim 10^{14}$ GeV). Taking the working assumption that three
neutrino masses are nearly degenerate, we calculate radiative
corrections to $V_0$ and $V'_0$ in their evolution down to the
electroweak scale ($\sim 10^2$ GeV). Three mixing angles of $V_0$
or $V'_0$ are essentially stable against radiative corrections in
the standard model (SM). In the minimal supersymmetric standard
model (MSSM), however, $V_0$ is in general disfavored and $V'_0$
can be compatible with current neutrino oscillation data if its
two Majorana phases $\alpha^{}_1$ and $\alpha^{}_2$ are properly
fine-tuned. We also find that it is possible to radiatively
generate the CP-violating phase $\delta$ from $\alpha^{}_1$ and
$\alpha^{}_2$, and $\delta$ may keep on staying at its quasi-fixed
point in either the SM or the MSSM.
\end{abstract}


\newpage

\framebox{\large\bf 1} ~
Current solar \cite{SNO}, atmospheric
\cite{SK}, reactor \cite{KM} and accelerator \cite{K2K} neutrino
experiments have provided us with very convincing evidence for the
existence of neutrino oscillations, a quantum phenomenon which can
naturally occur if neutrinos are massive and lepton flavors are
mixed. The property of lepton flavor mixing can be described by a
$3\times 3$ unitary matrix $V$. A parametrization of $V$,
advocated by the Particle Data Group \cite{PDG}, reads as
\begin{equation}
V = \left( \matrix{ c^{}_{12}c^{}_{13} & s^{}_{12}c ^{}_{13} &
s^{}_{13} e^{-i\delta} \cr -s^{}_{12}c^{}_{23}
-c^{}_{12}s^{}_{23}s^{}_{13} e^{i\delta} & c^{}_{12}c^{}_{23}
-s^{}_{12}s^{}_{23}s^{}_{13} e^{i\delta} & s^{}_{23}c^{}_{13} \cr
s^{}_{12}s^{}_{23} -c^{}_{12}c^{}_{23}s^{}_{13} e^{i\delta} &
-c^{}_{12}s^{}_{23} -s^{}_{12}c^{}_{23}s^{}_{13} e^{i\delta} &
c^{}_{23}c^{}_{13} } \right) \left ( \matrix{e^{i\alpha^{}_1/2} &
0 & 0 \cr 0 & e^{i\alpha^{}_2/2} & 0 \cr 0 & 0 & 1 \cr} \right )
\; ,
\end{equation}
where $c^{}_{ij} \equiv \cos\theta_{ij}$ and $s^{}_{ij} \equiv
\sin\theta_{ij}$ (for $ij=12,23$ and $13$). The phase parameters
$\alpha^{}_1$ and $\alpha^{}_2$ are usually referred as to the
Majorana CP-violating phases, because they have nothing to do with
CP or T violation in the neutrino-neutrino and
antineutrino-antineutrino oscillations. A global analysis of the
present experimental data yields \cite{Vissani} $30^\circ \leq
\theta_{12} \leq 38^\circ$, $36^\circ \leq \theta_{23} \leq
54^\circ$ and $0^\circ \leq \theta_{13} \leq 10^\circ$ as well as
$\Delta m^2_{21} \equiv m^2_2 - m^2_1 = (7.2 \cdot\cdot\cdot 8.9)
\times 10^{-5} ~{\rm eV}^2$ and $\Delta m^2_{32} \equiv m^2_3 -
m^2_2 = \pm (1.7 \cdot\cdot\cdot 3.3) \times 10^{-3} ~{\rm eV}^2$
at the $99\%$ confidence level. In contrast, three phases of $V$
are entirely unrestricted. A variety of new neutrino experiments
are underway, not only to detect the smallest flavor mixing angle
$\theta_{13}$ and the phase parameter $\delta$, but also to
constrain the Majorana phases $\alpha^{}_1$ and $\alpha^{}_2$.

To interpret the observed neutrino mass spectrum and the observed
bi-large neutrino mixing pattern, many theoretical and
phenomenological models have been proposed and discussed
\cite{Review}. A category of models or ans$\rm\ddot{a}$tze have
attracted some particular attention, because they can give rise to
the so-called tri-bimaximal neutrino mixing pattern \cite{HPS}:
\begin{equation}
V_0 \; =\; \left ( \matrix{ \sqrt{6}/3 & \sqrt{3}/3 & 0 \cr
-\sqrt{6}/6 & \sqrt{3}/3 & \sqrt{2}/2 \cr \sqrt{6}/6 & -\sqrt{3}/3
& \sqrt{2}/2 \cr} \right ) \; .
\end{equation}
Comparing between Eqs. (1) and (2), one may immediately observe
that $V_0$ has $\theta_{12} \approx 35.3^\circ$, $\theta_{23} =
45^\circ$, $\theta_{13} = 0^\circ$ and $\alpha^{}_1 = \alpha^{}_2
= 0^\circ$. The phase parameter $\delta$ is not well-defined in
$V_0$, as a consequence of $\theta_{13} = 0^\circ$. The results
$\sin^2 2\theta_{12} = 8/9$ and $\sin^2 2\theta_{23} =1$ are in
good agreement with current data on solar and atmospheric neutrino
oscillations. It is straightforward to generalize $V_0$ in order
to include two arbitrary Majorana phases,
\begin{equation}
V'_0 \; =\; \left ( \matrix{ \sqrt{6}/3 & \sqrt{3}/3 & 0 \cr
-\sqrt{6}/6 & \sqrt{3}/3 & \sqrt{2}/2 \cr \sqrt{6}/6 & -\sqrt{3}/3 &
\sqrt{2}/2 \cr} \right ) \left ( \matrix{e^{i\alpha^{}_1/2 } & 0 & 0
\cr 0 & e^{i\alpha^{}_2/2} & 0 \cr 0 & 0 & 1 \cr} \right ) \; .
\end{equation}
Although $V_0$ and $V'_0$ have the same impact on neutrino
oscillations, their consequences on the neutrinoless double-beta
decay are certainly different. In this sense, we refer to $V'_0$
as the {\it generalized} tri-bimaximal neutrino mixing pattern.

Such a special neutrino mixing pattern is in general expected to
result from an underlying flavor symmetry (e.g., the discrete
non-Abelian symmetry $A_4$ \cite{Ma,A4,He,Zee}) and its
spontaneous or explicit breaking. The latter is always necessary,
because a flavor symmetry itself cannot reproduce the observed
lepton mass spectra and predict the realistic lepton mixing
pattern simultaneously \cite{Grimus}. Specific and compelling
constructions of this kind of flavor symmetry breaking are a real
challenge and have been lacking, although some attempts have been
made in the literature \cite{Review}. The energy scale, at which a
proper flavor symmetry can be realized, may be considerably higher
than the electroweak scale ($\Lambda_{\rm EW} \sim 10^2$ GeV).
This new physics (NP) scale $\Lambda_{\rm NP}$ has actually been
identified with other known scales in some model-building works
\cite{Review}, including the grand-unification-theory scale
($\Lambda_{\rm GUT} \sim 10^{16}$ GeV) or the seesaw scale
($\Lambda_{\rm SS} \sim 10^{14}$ GeV). In this case, radiative
corrections to the relevant model parameters between $\Lambda_{\rm
EW}$ and $\Lambda_{\rm NP}$ must be taken into account \cite{RGE}.

One may then ask whether the generalized tri-bimaximal neutrino
mixing pattern is stable or not against radiative corrections, if
it is derived from an underlying (broken) flavor symmetry within
an unspecified mechanism at $\Lambda_{\rm NP}$ ($\gg \Lambda_{\rm
EW}$). The main purpose of this paper is just to answer this
question by considering both the standard model (SM) and its
minimal supersymmetric extension (MSSM) below $\Lambda_{\rm NP}$.
The only effective dimension-5 operator of light Majorana
neutrinos reads as
\begin{equation}
{\cal L}_\nu \; =\; \frac{1}{2} \kappa^{}_{ij} \left (H \cdot L_i
\right ) \left ( H \cdot L_j \right ) + {\rm h.c.} \; ,
\end{equation}
where $H$ denotes the SM Higgs (or the MSSM Higgs with the
appropriate hypercharge), $L_i$ (for $i=1,2,3$) stand for the
leptonic $SU(2)_{\rm L}$ doublets, and $\kappa$ is a symmetric
neutrino coupling matrix. After spontaneous gauge symmetry
breaking at $\Lambda_{\rm EW}$, we arrive at the effective
neutrino mass matrix $M_\nu = v^2\kappa$ (SM) or $M_\nu =
v^2\kappa \sin^2\beta$ (MSSM), where $v \approx 174$ GeV and
$\tan\beta$ is the ratio of the vacuum expectation values of two
Higgs fields in the MSSM. Between $\Lambda_{\rm EW}$ and
$\Lambda_{\rm NP}$, the most important radiative correction to
$\kappa$ is proportional to $\ln (\Lambda_{\rm NP}/\Lambda_{\rm
EW})$ and can be evaluated by using the one-loop renormalization
group equations (RGEs) \cite{RGE}. It is then possible to
calculate the RGE effects on the lepton flavor mixing parameters
analytically and numerically.

In the working assumption that three neutrino masses are nearly
degenerate, we are going to calculate radiative corrections to
$V_0$ and $V'_0$. We show that both $V_0$ and $V'_0$ are stable
against radiative corrections in the SM, but only $V'_0$ with the
proper fine-tuning of $(\alpha^{}_2 -\alpha^{}_1)$ is allowed in
the MSSM. In addition, the CP-violating parameter $\delta$ can be
radiatively generated from $\alpha^{}_1$ and $\alpha^{}_2$. A
peculiar feature of $\delta$ is that it may keep on staying at its
quasi-fixed point in both the SM and the MSSM.

\vspace{0.2cm}

\framebox{\large\bf 2} ~
Taking account of the seesaw mechanism
\cite{SS} as a natural idea to understand the origin of neutrino
masses and lepton flavor mixing, we assume the new physics (i.e.,
new flavor symmetry) scale $\Lambda_{\rm NP}$ is close to the
seesaw scale $\Lambda_{\rm SS} \sim 10^{14}$ GeV. Below
$\Lambda_{\rm NP}$, the effective neutrino coupling matrix
$\kappa$ obeys the one-loop RGE \cite{RGE1}
\footnote{Note that $\Lambda_{\rm NP} \sim \Lambda_{\rm SS}$ is an
effective working assumption, in which the possible mass hierarchy
of three heavy right-handed neutrinos $N_i$ (for $i=1,2,3$) is
omitted. If $\Lambda_{\rm NP} \sim \Lambda_{\rm GUT}$ ($>
\Lambda_{\rm SS}$) is assumed and the mass hierarchy of $N_i$ is
considered, then very strong seesaw threshold effects may appear
in the RGE evolution of relevant model parameters (see Ref.
\cite{Threshold} for detailed discussions).}:
\begin{equation}
16\pi^2 \frac{{\rm d}\kappa}{{\rm d}t} = \alpha \kappa + C \left [
\left (Y^{}_lY^\dagger_l \right ) \kappa + \kappa \left (Y^{}_l
Y^\dagger_l \right )^T \right ] \; ,
\end{equation}
in which $t\equiv \ln (\mu/\Lambda_{\rm NP})$ with $\mu$ being an
arbitrary renormalization scale below $\Lambda_{\rm NP}$ but above
$\Lambda_{\rm EW}$. In the SM, $C=-1.5$ and $\alpha \approx
-3g^2_2 + 6y^2_t + \lambda$; and in the MSSM, $C=1$ and $\alpha
\approx -1.2g^2_1 - 6g^2_2 + 6 y^2_t$, where $g^{}_1$ and $g^{}_2$
denote the gauge couplings, $y^{}_t$ stands for the top-quark
Yukawa coupling, and $\lambda$ represents the Higgs self-coupling
in the SM \cite{RGE1}. In the flavor basis where the
charged-lepton Yukawa coupling matrix is diagonal and real
(positive), we have $\kappa = V \overline{\kappa} V^T$ with
$\overline{\kappa} = {\rm Diag}\{\kappa^{}_1, \kappa^{}_2,
\kappa^{}_3\}$. The neutrino masses at $\Lambda_{\rm EW}$ are
given by $m^{}_i = v^2 \kappa^{}_i$ (SM) or $m^{}_i = v^2
\kappa^{}_i \sin^2\beta$ (MSSM). One can then derive the RGEs for
$(\kappa^{}_1, \kappa^{}_2, \kappa^{}_3)$, $(\theta_{12},
\theta_{23}, \theta_{13})$ and $(\delta, \alpha^{}_1,
\alpha^{}_2)$ from Eq. (5), just like the previous works done in
Refs. \cite{RGE1,Threshold,Antusch,Luo}.

To be specific, we assume the masses of three Majorana neutrinos
are nearly degenerate; i.e., $m^{}_1 \approx m^{}_2 \approx
m^{}_3$. Such a working assumption makes sense at least for two
practical reasons: (1) it might hint at the slight breaking of an
exact $S(3)$ permutation symmetry \cite{FX96} or other possible
flavor symmetries, from which the tri-bimaximal neutrino mixing
pattern can naturally arise; and (2) more significant RGE running
effects on three mixing angles and three CP-violating phases can
manifest themselves in this interesting case. Furthermore, it is
helpful to make some analytical approximations for the results
obtained in Refs. \cite{RGE1,Threshold,Antusch,Luo} by taking
account of the smallness of $\sin\theta_{13}$ and $\Delta
m^2_{21}/\Delta m^2_{32}$. We arrive at
\begin{eqnarray}
\frac{{\rm d} m^{}_1}{{\rm d} t} & \approx &
\frac{m^{}_1}{16\pi^2} \left ( \alpha + 2 C y_{\tau}^2 s_{12}^2
s_{23}^2 \right ) \; ,
\nonumber \\
\frac{{\rm d} m^{}_2}{{\rm d} t} & \approx &
\frac{m^{}_2}{16\pi^2} \left ( \alpha + 2 C y_{\tau}^2 c_{12}^2
s_{23}^2 \right ) \; ,
\nonumber \\
\frac{{\rm d} m^{}_3}{{\rm d} t} & \approx &
\frac{m^{}_3}{16\pi^2} \left ( \alpha + 2 C y_{\tau }^2 c_{23}^2
\right ) \;
\end{eqnarray}
to a good degree of accuracy, where $y^{}_\tau$ denotes the
tau-lepton Yukawa coupling. Given the approximate degeneracy of
three neutrino masses, the RGEs of $(\theta_{12}, \theta_{23},
\theta_{13})$ and $(\delta, \alpha^{}_1, \alpha^{}_2)$ in Refs.
\cite{RGE1,Threshold,Antusch,Luo} are simplified to
\begin{eqnarray}
\frac{{\rm d} \theta^{}_{12}}{{\rm d} t} & \approx & -
\frac{Cy^2_\tau}{4\pi^2} \cdot \frac{m^2_1}{\Delta m^2_{21}} ~
c^{}_{12} s^{}_{12} s^2_{23} \cos^2 \frac{\alpha^{}_2 -
\alpha^{}_1}{2} \; ,
\nonumber \\
\frac{{\rm d} \theta^{}_{23}}{{\rm d} t} & \approx & -
\frac{Cy^2_\tau}{4\pi^2} \cdot \frac{m^2_1}{\Delta m^2_{32}} ~
c^{}_{23} s^{}_{23} \left ( c^2_{12} \cos^2 \frac{\alpha^{}_2}{2}
+ s^2_{12} \cos^2 \frac{\alpha^{}_1}{2} \right ) \; ,
\nonumber \\
\frac{{\rm d} \theta^{}_{13}}{{\rm d} t} & \approx & -
\frac{Cy^2_\tau}{8\pi^2} \cdot \frac{m^2_1}{\Delta m^2_{32}} ~
c^{}_{12} s^{}_{12} c^{}_{23} s^{}_{23} \left [ \cos \left (\delta
+\alpha^{}_2 \right ) - \cos \left (\delta + \alpha^{}_1 \right )
\right ] \; ;
\end{eqnarray}
as well as
\begin{eqnarray}
\frac{{\rm d} \delta}{{\rm d} t} ~ & \approx &
\frac{Cy^2_\tau}{8\pi^2} \left [ \frac{m^2_1}{\Delta m^2_{32}}
\cdot \frac{c^{}_{12} s^{}_{12} c^{}_{23} s^{}_{23}}{s^{}_{13}}
\left [ \sin \left ( \delta + \alpha^{}_2 \right ) - \sin \left (
\delta + \alpha^{}_1 \right ) + \chi \right ] +
\frac{m^2_1}{\Delta m^2_{21}} ~ s^2_{23} \sin \left ( \alpha^{}_2
- \alpha^{}_1 \right ) \right ] \; ,
\nonumber \\
\frac{{\rm d} \alpha^{}_1}{{\rm d} t} & \approx & -
\frac{Cy^2_\tau}{4\pi^2} \cdot \frac{m^2_1}{\Delta m^2_{21}} ~
c^2_{12} s^2_{23} \sin \left ( \alpha^{}_2 - \alpha^{}_1 \right )
\; ,
\nonumber \\
\frac{{\rm d} \alpha^{}_2}{{\rm d} t} & \approx & -
\frac{Cy^2_\tau}{4\pi^2} \cdot \frac{m^2_1}{\Delta m^2_{21}} ~
s^2_{12} s^2_{23} \sin \left ( \alpha^{}_2 - \alpha^{}_1 \right )
\; ,
\end{eqnarray}
where
\begin{equation}
\chi \; =\; \frac{\Delta m^2_{21}}{\Delta m^2_{32}} \left [ \sin
\left (\delta + \alpha^{}_1 \right ) + \sin \delta \right ] \; .
\end{equation}
Note that the $\chi$-term is not negligible only in the special
case that $\alpha^{}_1 \approx \alpha^{}_2$ holds and $s^{}_{13}$
is extremely small.
Some qualitative comments on Eqs. (7) and (8) are in order.

(a) The mixing angle $\theta_{12}$ is in general more sensitive to
radiative corrections than $\theta_{23}$ and $\theta_{13}$
\cite{RGE1,Antusch}. Given $\theta_{12} \approx 35.3^\circ$ as a
result of the tri-bimaximal neutrino mixing at $\Lambda_{\rm NP}$,
the RGE running effect has to be sufficiently suppressed such that
$\theta_{12}$ can finally run into the experimentally-allowed
range $30^\circ \leq \theta_{12} \leq 38^\circ$ at low energies.
This requirement is certainly satisfied in the SM with $m^{}_1
\sim {\cal O}(0.1 ~ {\rm eV})$, in which $\theta_{12}$ slightly
decreases in the RGE evolution from $\Lambda_{\rm NP}$ to
$\Lambda_{\rm EW}$. In the MSSM, however, $\theta_{12}$ must
evolve to a bigger value at $\Lambda_{\rm EW}$. Hence the
fine-tuning of $(\alpha^{}_2 -\alpha^{}_1)$ is necessary for large
values of $\tan\beta$, so as to keep the evolution effect of
$\theta_{12}$ insignificant \cite{RGE1,Antusch,Haba}. One can see
that two Majorana phases of $V'_0$ play a very non-trivial role in
the calculation of radiative corrections. In other words, the
tri-bimaximal neutrino mixing pattern $V_0$ and its generalized
counterpart $V'_0$ are distinguishable in model building by taking
into account their different RGE running behaviors.

(b) Different from $\theta_{12}$, the mixing angles $\theta_{23}$
and $\theta_{13}$ are expected to be less sensitive to radiative
corrections. Hence $\theta_{23}$ at $\Lambda_{\rm EW}$ may
slightly deviate from its initial value $\theta_{23} =45^\circ$ at
$\Lambda_{\rm NP}$ as a consequence of the RGE running. On the
other hand, $\theta_{13}$ can be radiatively generated, although
its value at $\Lambda_{\rm EW}$ must be rather small. Note that
the tri-bimaximal neutrino mixing pattern $V_0$ keeps
CP-conserving in the RGE evolution from $\Lambda_{\rm NP}$ to
$\Lambda_{\rm EW}$. As for the generalized tri-bimaximal neutrino
mixing scenario $V'_0$, it is possible to generate both the mixing
angle $\theta_{13}$ and the CP-violating phase $\delta$
radiatively
\footnote{For a more generic study of this problem, we refer
readers to the works done by Casas {\it et al} in Ref.
\cite{RGE1}, Antusch {\it et al} in Ref. \cite{Antusch}, and Luo
{\it et al} in Ref. \cite{Luo}.}.
The latter results from two non-trivial Majorana phases (or one of
them) in the RGE of $\delta$. This observation implies that the
RGE-corrected $V'_0$ may give rise to a non-vanishing Jarlskog
invariant \cite{J} at $\Lambda_{\rm EW}$, leading to observable CP
violation in neutrino oscillations.

(c) It should be noted that $\delta$ is not well-defined at
$\Lambda_{\rm NP}$, where $\theta_{13}$ is exactly vanishing in
either $V_0$ or $V'_0$. This point can clearly be seen from Eq.
(8), in which the derivative of $\delta$ diverges in the
$\theta_{13} \rightarrow 0$ limit. Nevertheless, it has been shown
in Ref. \cite{Antusch} that there exists an analytic continuation
of $\delta$, such that it remains well-defined even when
$\theta_{13}$ approaches zero. Hence $\theta_{13}$ and $\delta$
can simultaneously be generated from the RGE running effects in
the generalized tri-bimaximal neutrino mixing scenario. A peculiar
RGE behavior of $\delta$ is that it can keep on staying at its
quasi-fixed point just below $\Lambda_{\rm NP}$, as one can see in
the subsequent numerical examples.

\vspace{0.2cm}

\framebox{\large\bf 3} ~
Now let us quantitatively illustrate
radiative corrections to $V_0$ and $V'_0$ by taking a few
numerical examples
\footnote{Our numerical calculations follow a ``running and
diagonalizing" procedure \cite{Antusch}: we first compute the RGE
evolution of lepton mass matrices and then extract their mass
eigenvalues and flavor mixing parameters at $\Lambda_{\rm EW}$.
Because $\theta_{13} =0$ holds exactly at $\Lambda_{\rm NP}$ and
$\delta$ is always associated with $s^{}_{13}$ in the chosen
parametrization of $V$, any initial input of $\delta$ is allowed
but it does not take any effect in the RGE running. The finite
running result of $\delta$ is actually attributed to the initial
values of two Majorana phases $\alpha^{}_1$ and $\alpha^{}_2$.}.
The eigenvalues of $Y_l$ at $\Lambda_{\rm NP}$ are chosen in such
a way that they can correctly run to their low-energy values. We
typically take $m^{}_1 \sim 0.2$ eV, which is consistent with the
working assumption $m^{}_1 \approx m^{}_2 \approx m^{}_3$ made
above. The initial values of three mixing angles at $\Lambda_{\rm
NP}$ are fixed: $\theta_{12} \approx 35.3^\circ$, $\theta_{23} =
45^\circ$ and $\theta_{13} =0^\circ$, predicted by $V_0$ or
$V'_0$. Two unknown Majorana phases in $V'_0$ are adjustable in
our numerical calculations, in which we choose $m^{}_H = 140$ GeV
(SM) or $\tan\beta = 10$ (MSSM) as a typical and instructive
input. The primary results are shown in Tables I and II together
with Figs. 1 and 2. Some discussions are in order.

{\it (A) In the SM}. ~ It is demonstrated that the RGE running
effects on three mixing angles ($\theta_{12}$, $\theta_{23}$ and
$\theta_{13}$) are small enough in the SM, thus either $V_0$ or
$V'_0$ at $\Lambda_{\rm NP}$ can agree with current neutrino
oscillation data at low energies. For $V_0$, CP conservation keeps
to hold in the RGE evolution from $\Lambda_{\rm NP}$ to
$\Lambda_{\rm EW}$ (see Case I in Table I). But for $V'_0$, the
radiative generation of $\delta$ can always take place, only if
$\alpha^{}_1$ or $\alpha^{}_2$ is initially non-vanishing (see
Cases II and III in Table I). Because of $\theta_{13} =0^\circ$ at
$\Lambda_{\rm NP}$, an extraordinarily large RGE correction to the
CP-violating phase $\delta$ arises from the term proportional to
$1/s^{}_{13}$ on the right-hand side of Eq. (8). It turns out that
$\delta$ keeps on staying at its {\it quasi-fixed point} (see Fig.
1 for illustration). In contrast, the running of $\alpha^{}_1$ and
$\alpha^{}_2$ is so tiny that they are essentially stable between
$\Lambda_{\rm NP}$ and $\Lambda_{\rm EW}$.

We find that the analytical approximation made in Eq. (8) together
with Eq. (9) is helpful to understand the quasi-fixed point of the
CP-violating phase $\delta$ in its RGE evolution from
$\Lambda_{\rm NP}$ to $\Lambda_{\rm EW}$. At such a quasi-fixed
point, the condition ${\rm d} \delta/{\rm d}t \approx 0$ should be
satisfied. This observation essentially implies that either
\begin{equation}
\sin \left ( \hat{\delta} + \hat{\alpha}^{}_2 \right ) - \sin
\left ( \hat{\delta} + \hat{\alpha}^{}_1 \right ) \; \approx \; 0
\;
\end{equation}
with $\hat{\alpha}^{}_1 \neq \hat{\alpha}^{}_2$ or
\begin{equation}
\sin \left ( \hat{\delta} + \hat{\alpha}^{}_1 \right ) + \sin
\hat{\delta} \; \approx \; 0 \;
\end{equation}
with $\hat{\alpha}^{}_1 \approx \hat{\alpha}^{}_2$ should hold in
the leading-order approximation; i.e., the $1/s^{}_{13}$ term in
the expression of ${\rm d} \delta/{\rm d}t$ must be strongly
suppressed around the quasi-fixed point, where the values of three
CP-violating phases are denoted by $\hat{\delta}$,
$\hat{\alpha}^{}_1$ and $\hat{\alpha}^{}_2$. The simple solutions
to Eqs. (10) and (11) are
\begin{equation}
\hat{\delta} \; \approx \; - \frac{1}{2} \left ( \hat{\alpha}^{}_1
+ \hat{\alpha}^{}_2 \right ) + \left (n + \frac{1}{2} \right ) \pi
\; , \;\;\;\;\; (\hat{\alpha}^{}_1 \neq \hat{\alpha}^{}_2) \;
\end{equation}
and
\begin{equation}
\hat{\delta} \; \approx \; - \frac{\hat{\alpha}^{}_1}{2} + n \pi
\; , \;\;\;\;\; (\hat{\alpha}^{}_1 \approx \hat{\alpha}^{}_2) \;
\end{equation}
(for $n = 0, \pm 1, \pm 2, \cdot\cdot\cdot$), respectively. These
two possibilities correspond to the numerical examples given in
Cases II and III in Table I or Fig. 1(a) and (b).

The tiny magnitude of $\theta_{13}$ at $\Lambda_{\rm EW}$ implies
that it is easy to rule out $V_0$ or $V'_0$ at $\Lambda_{\rm NP}$,
after $\theta_{13} \neq 0^\circ$ is experimentally established.
Indeed, the sensitivity of a few currently-proposed reactor
neutrino experiments to $\theta_{13}$ is at the level of
$\theta_{13} \sim 3^\circ$ or $\sin^2 2\theta_{13} \sim 0.01$
\cite{Wang}. Since $\theta_{13}$ is considerably suppressed, as
shown in our numerical examples, it will be extremely difficult to
measure $\delta$ (even if $\delta \sim \pm 90^\circ$) in any
long-baseline neutrino oscillation experiments. In this case, only
the neutrinoless double-beta decay could be used to distinguish
$V'_0$ from $V_0$, because its effective mass $\langle
m\rangle_{ee}$ is sensitive to the Majorana phases
\footnote{We have $\langle m\rangle_{ee} = \left | m^{}_1 c^2_{12}
c^2_{13} e^{i\alpha^{}_1} + m^{}_2 s^2_{12} c^2_{13}
e^{i\alpha^{}_2} + m^{}_3 s^2_{13} e^{-2i\delta} \right |$ by
using the parametrization of $V$ given in Eq. (1). This result
implies that it is actually ill to refer to $\delta$ as the Dirac
phase in this ``standard" parametrization. A different phase
convention of $V$ has been proposed in Ref. \cite{FX01} to forbid
$\delta$ to appear in $\langle m\rangle_{ee}$.}.
We know that $V_0$ predicts $\langle m\rangle_{ee} \approx m^{}_1$
and $V'_0$ yields
\begin{equation}
\langle m\rangle'_{ee} \; \approx \; m^{}_1 \sqrt{1 - \sin^2
2\theta_{12} \sin^2 \frac{\alpha^{}_2 - \alpha^{}_1}{2}} \;
\end{equation}
in the assumption that three neutrino masses are nearly
degenerate. The ratio of $\langle m\rangle'_{ee}$ to $\langle
m\rangle_{ee}$ may take its minimal value $\left [\langle
m\rangle'_{ee}/\langle m\rangle_{ee} \right ]^{}_{\rm min} \approx
\cos 2\theta_{12} \approx 0.34$ (for $\theta_{12} \approx
35^\circ$), when $\alpha^{}_2 -\alpha^{}_1 \approx \pm \pi$ is
satisfied.

We remark that the RGE running behaviors of $V_0$ and $V'_0$ are
quite different in the SM. This difference can definitely affect
the model building in understanding the origin of $V_0$ or $V'_0$.
It is worth mentioning that a particular mass model of charged
leptons and neutrinos with the non-Abelian symmetry $A_4$, from
which $V'_0$ can be derived at a superhigh energy scale, has been
proposed and discussed in Ref. \cite{He}. Its low-energy
consequences are actually within the scope of our generic RGE
analysis.

{\it (B) In the MSSM}. ~
We have pointed out that the mixing angle
$\theta_{12}$ is quite sensitive to radiative corrections, and it
always runs to a bigger value at $\Lambda_{\rm EW}$ from the
initial value $\theta_{12} \approx 35.3^\circ$ at $\Lambda_{\rm
NP}$ in the MSSM. Considering the tri-bimaximal neutrino mixing
pattern $V_0$ and taking $\tan\beta =10$ as a typical input, we
find that the RGE running result of $\theta_{12}$ at $\Lambda_{\rm
EW}$ exceeds its experimental upper bound (i.e., $\theta_{12} \leq
38^\circ$ at the $99\%$ confidence level \cite{Vissani}). We can
therefore conclude that a neutrino mass model predicting $V_0$ at
$\Lambda_{\rm NP}$ is in general disfavored in the MSSM.

This situation will change, if the generalized tri-bimaximal
neutrino mixing pattern $V'_0$ is concerned. The reason is simply
that the increase of $\theta_{12}$ during its RGE running can be
controlled by the Majorana phase factor $\cos^2 (\alpha^{}_2
-\alpha^{}_1)$, as shown in Eq. (7). Hence the proper fine-tuning
of $(\alpha^{}_2 -\alpha^{}_1)$ will allow $\theta_{12}$ to mildly
evolve into its experimental range $30^\circ \leq \theta_{12} \leq
38^\circ$ at $\Lambda_{\rm EW}$. Given $\tan\beta =10$, an
approximate bound on $(\alpha^{}_2 -\alpha^{}_1)$ is found to be
$154^\circ \leq |\alpha^{}_2 -\alpha^{}_1| \leq 206^\circ$ in our
calculation. We present an explicit numerical example in Table II
and Fig. 2, just for the purpose of illustration.

As a direct consequence of $\theta_{13} =0^\circ$ at $\Lambda_{\rm
NP}$, a very significant RGE correction to the CP-violating phase
$\delta$ arises from the term proportional to $1/s^{}_{13}$ in
${\rm d}\delta/{\rm d}t$. Thus $\delta$ can keep on staying at its
{\it quasi-fixed point} in the MSSM, just like the case in the SM.
Following the discussions given above, one may approximately
arrive at the relations given in Eqs. (12) and (13) at the
quasi-fixed point. The latter possibility has been ruled out by
taking into account the evolution of $\theta_{12}$, and the former
possibility is illustrated in Table II or Fig. 2, where the rather
mild running behaviors of $\alpha^{}_1$ and $\alpha^{}_2$ can also
be seen. In this example, $\theta_{13} \approx 1.4^\circ$ together
with $\delta \approx -78^\circ$ can be radiatively generated at
$\Lambda_{\rm EW}$. This result implies that the magnitude of the
Jarlskog invariant (i.e., ${\cal J} = s^{}_{12} c^{}_{12}
s^{}_{23} c^{}_{23} s^{}_{13} c^2_{13} \sin\delta$) can be as
large as about $0.6\%$, probably leading to the observable
CP-violating effect in the future long-baseline neutrino
oscillation experiments.

Let us stress that the quasi-fixed point in the RGE evolution of
$\delta$ is in general unavoidable for those neutrino mixing
patterns with $\theta_{13} =0$. The above analytic understanding
of such a quasi-fixed point is new and helpful for specific model
building. If the phase convention of $V$ in Eq. (1) is replaced by
that proposed in \cite{FX01}, its corresponding Majorana phases
$\rho = \delta + \alpha^{}_1/2$ and $\sigma = \delta +
\alpha^{}_2/2$ will also have the quasi-fixed points in their RGE
evolution. This point can easily be understood with the help of
Eq. (8): the dominant term of ${\rm d}\delta/{\rm d}t$
(proportional to $1/s^{}_{13}$) will enter ${\rm d}\rho/{\rm d}t$
and ${\rm d}\sigma/{\rm d}t$, such that the RGE running behaviors
of $\delta$, $\rho$ and $\sigma$ are essentially identical
\cite{Luo}. In short, the ``standard" parametrization of $V$ taken
in Eq. (1) is more convenient in discussing the issue of
quasi-fixed points, while the phase convention of $V$ advocated in
Ref. \cite{FX01} is more convenient in discussing the neutrinoless
double-beta decay (i.e., $\langle m\rangle_{ee}$ is dependent on
$\rho$ and $\sigma$ but independent of $\delta$).

\vspace{0.2cm}

\framebox{\large\bf 4} ~ We have argued that the tri-bimaximal
neutrino mixing pattern $V_0$ or its generalized counterpart
$V'_0$ is very likely to result from an underlying flavor
symmetry, and this new symmetry is most likely to be realized at a
superhigh energy scale. Supposing that this new physics scale is
close to the neutrino seesaw scale ($\sim 10^{14}$ GeV), we have
calculated the one-loop RGE effects on $V_0$ and $V'_0$ in their
evolution down to the electroweak scale ($\sim 10^2$ GeV) in the
working assumption that three neutrino masses are nearly
degenerate. It is found that three mixing angles of $V_0$ or
$V'_0$ are essentially insensitive to radiative corrections in the
SM. In the MSSM, however, $V_0$ is in general disfavored and
$V'_0$ can be compatible with current neutrino oscillation data if
its two Majorana phases $\alpha^{}_1$ and $\alpha^{}_2$ are
properly fine-tuned. We have also shown that it is possible to
radiatively generate the CP-violating phase $\delta$ from
$\alpha^{}_1$ and $\alpha^{}_2$, and $\delta$ may keep on staying
at its quasi-fixed point in both the SM and the MSSM.

Although a detailed RGE analysis of the tri-bimaximal neutrino
mixing scenario has not been done before, some of our results have
actually been observed by some other authors in their analyses of
radiative corrections to the neutrino mass spectrum and realistic
lepton flavor mixing patterns, whose forms are more or less
similar to the tri-bimaximal neutrino mixing pattern $V_0$.
However, our present work is new in two important aspects: (1) we
generalize $V_0$ to $V'_0$ with two arbitrary Majorana phases,
because the latter is more interesting and can naturally appear in
some specific neutrino mass models; (2) we explore the quasi-fixed
point in the RGE evolution of $\delta$ and present an analytic
understanding of this non-trivial phenomenon.

The afore-mentioned RGE running behaviors of $V_0$ and $V'_0$ are
expected to be useful for model building at a superhigh energy
scale. A similar study can be extended to some other interesting
neutrino mixing patterns. For example, the pattern
\begin{equation}
U_0 \; =\; \left ( \matrix{ \sqrt{3}/2 & 1/2 & 0 \cr -\sqrt{2}/4 &
\sqrt{6}/4 & \sqrt{2}/2 \cr \sqrt{2}/4 & -\sqrt{6}/4 & \sqrt{2}/2
\cr} \right ) \left ( \matrix{e^{i\alpha^{}_1/2}  & 0 & 0 \cr 0 &
e^{i\alpha^{}_2/2} & 0 \cr 0 & 0 & 1 \cr} \right ) \; ,
\end{equation}
which has $\theta_{12} = 30^\circ$, $\theta_{23} = 45^\circ$ and
$\theta_{13} = 0^\circ$ \cite{Xing03}, is rather analogous to the
generalized tri-bimaximal neutrino mixing pattern $V'_0$. If $U_0$
is derived from an underlying flavor symmetry at an energy scale
close to the seesaw scale, then its sensitivity to radiative
corrections must be very similar to that of $V'_0$.

To examine whether such a special lepton mixing scenario is viable
or not in a high-energy neutrino mass model, it is crucial to
measure the smallest mixing angle $\theta_{13}$ and the
CP-violating phase $\delta$ in the future neutrino oscillation
experiments. Furthermore, any experimental information about the
Majorana phases $\alpha^{}_1$ and $\alpha^{}_2$ is welcome and
extremely important, in order to distinguish one model from
another through their different sensitivities to radiative
corrections.


{\it Acknowledgments:} ~ We would like to thank J.W. Mei for his
nice {\it Mathematica} program and H. Zhang for very useful
discussions. This work was supported in part by the National
Natural Science Foundation of China.

\newpage

\begin{table}
\caption{Radiative corrections to $V_0$ (Case I) and $V'_0$ (Cases
II and III) from $\Lambda_{\rm NP} \sim 10^{14}$ GeV to
$\Lambda_{\rm EW} \sim 10^2$ GeV in the SM. The Higgs mass $m^{}_H
= 140$ GeV has typically been input in our numerical calculation.
Note that $\delta$ is not well-defined in the $\theta_{13} =0$
limit at $\Lambda_{\rm NP}$, but its running behavior is
independent of this ambiguity and is fixed by the initial values
of $\alpha^{}_1$ and $\alpha^{}_2$.}
\vspace{0.2cm}
\begin{center}
\begin{tabular}{c|cc|cc|cc}
Parameter & \multicolumn{2}{c}{Case I ($V_0$)}
& \multicolumn{2}{c}{Case II ($V'_0$)}
& \multicolumn{2}{c}{Case III ($V'_0$)}  \\
& $\Lambda_{\rm NP}$ & $\Lambda_{\rm EW}$ & $\Lambda_{\rm NP}$ &
$\Lambda_{\rm EW}$
& $\Lambda_{\rm NP}$ & $\Lambda_{\rm EW}$ \\
\hline
$m^{}_1 ({\rm eV} )$ & 0.310 & 0.200 & 0.310 & 0.200 & 0.310 & 0.200 \\
$\Delta m^2_{21} ( 10^{-5} ~{\rm eV}^2 )$ & 18.83 & 7.91 & 18.83 & 7.91 & 18.83 & 7.91 \\
$\Delta m^2_{31} ( 10^{-3} ~{\rm eV}^2 )$ & 5.31 & 2.21 & 5.31 & 2.21 & 5.31 & 2.21 \\
\hline
$\theta_{12}$ & $35.26^\circ$ & $34.48^\circ$ & $35.26^\circ$
& $35.24^\circ$ & $35.26^\circ$ & $34.48^\circ$ \\
$\theta_{23}$ & $45.0^\circ$ & $44.94^\circ$ & $45.0^\circ$
& $44.97^\circ$ & $45.0^\circ$ & $44.96^\circ$ \\
$\theta_{13}$ & $0^\circ$ & $0.001^\circ$ & $0^\circ$ &
$0.0288^\circ$ & $0^\circ$ & $0.0008^\circ$ \\
\hline
$\delta$ & --- & $0^\circ$ & --- & $90.61^\circ$
& --- & $140.0^\circ$ \\
$\alpha^{}_1$ & $0^\circ$ & $0^\circ$ & $260.0^\circ$ &
$260.38^\circ$
& $80.0^\circ$ & $80.0^\circ$ \\
$\alpha^{}_2$ & $0^\circ$ & $0^\circ$ & $100.0^\circ$ &
$100.19^\circ$
& $80.0^\circ$ & $80.0^\circ$ \\
\end{tabular}
\end{center}
\end{table}

\begin{table}
\caption{Radiative corrections to $V'_0$ from $\Lambda_{\rm NP}
\sim 10^{14}$ GeV to $\Lambda_{\rm EW} \sim 10^2$ GeV in the MSSM.
In our numerical calculation, $\tan\beta = 10$ has typically been
input. Note that $\delta$ is not well-defined in the $\theta_{13}
=0$ limit at $\Lambda_{\rm NP}$, but its running behavior is
independent of this ambiguity and is fixed by the initial values
of $\alpha^{}_1$ and $\alpha^{}_2$.} \vspace{0.2cm}
\begin{center}
\begin{tabular}{c|cc}
Parameter & Input at $\Lambda_{\rm NP}$ & Output at $\Lambda_{\rm EW}$ \\
\hline
$m^{}_1 ({\rm eV} )$ & 0.241 & 0.201   \\
$\Delta m^2_{21} ( 10^{-5} ~{\rm eV}^2 )$ & 17.0 & 8.19   \\
$\Delta m^2_{31} ( 10^{-3} ~{\rm eV}^2 )$ & 3.3 & 2.21   \\
\hline
$\theta_{12}$ & $35.26^\circ$ & $36.38^\circ$  \\
$\theta_{23}$ & $45.0^\circ$ & $46.22^\circ$  \\
$\theta_{13}$ & $0^\circ$ & $1.367^\circ$  \\
\hline
$\delta$ & --- & $-77.85^\circ$  \\
$\alpha^{}_1$ & $260.0^\circ$ & $245.17^\circ$  \\
$\alpha^{}_2$ & $100.0^\circ$ & $92.27^\circ$  \\
\end{tabular}
\end{center}
\end{table}

\newpage

\begin{figure}
\begin{center}
\vspace{-1cm}
\includegraphics[width=15.5cm,height=22.5cm]{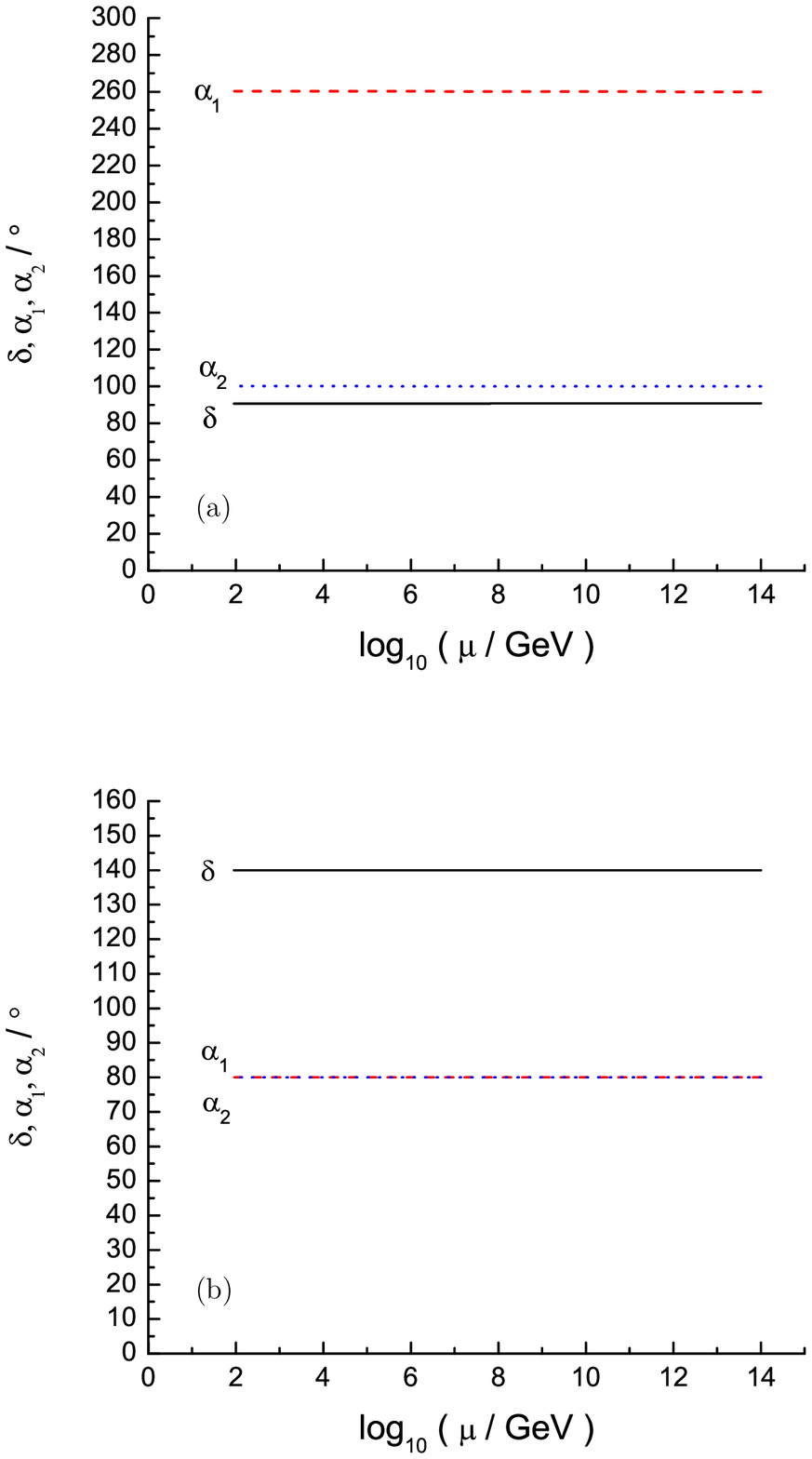}
\vspace{-4.cm} \caption{The RGE running behaviors of three
CP-violating phases of $V'_0$ from $\Lambda_{\rm NP}$ to
$\Lambda_{\rm EW}$ in the SM. The input and output values of other
relevant parameters can be found from Table I.}
\end{center}
\end{figure}

\newpage

\begin{figure}
\begin{center}
\vspace{8cm}
\includegraphics[width=15.5cm,height=22.5cm]{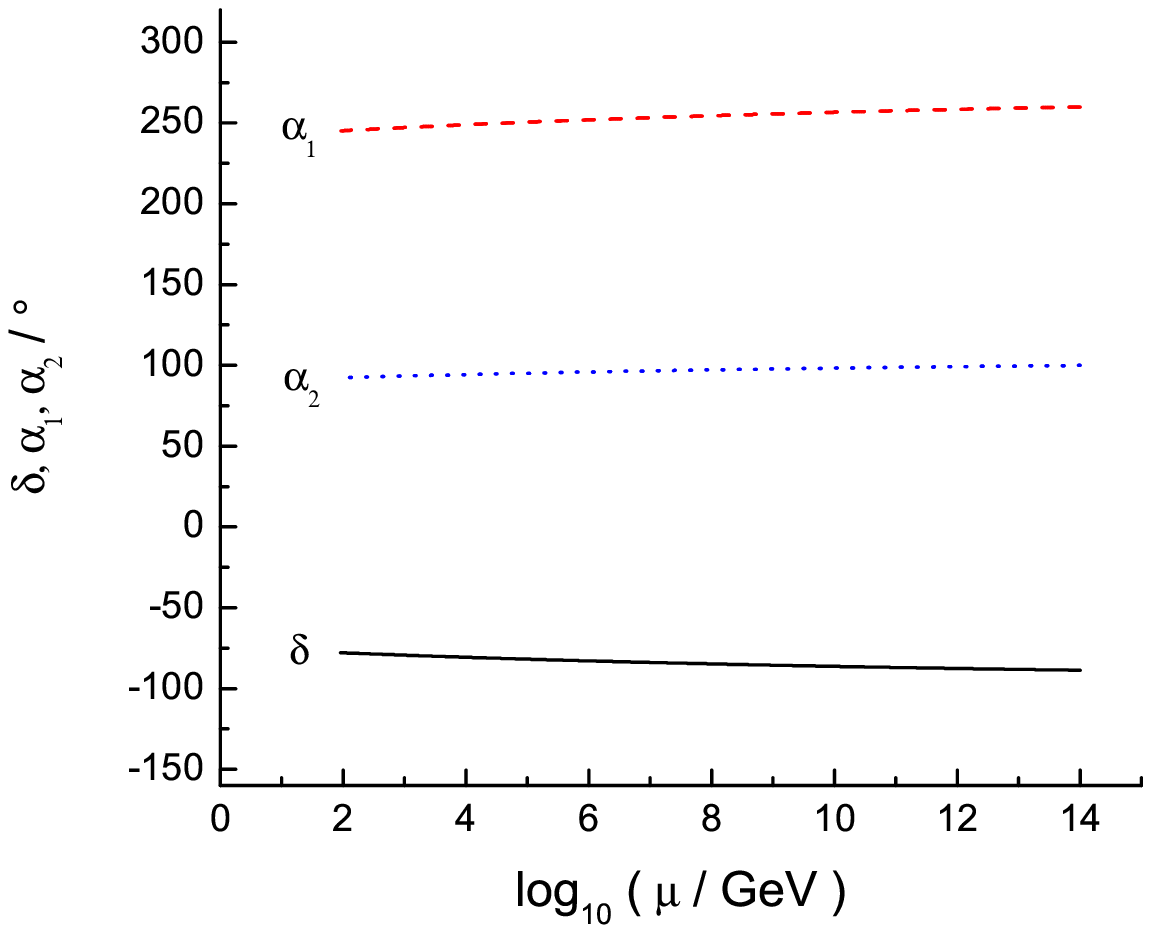}
\vspace{-12.5cm} \caption{The RGE running behaviors of three
CP-violating phases of $V'_0$ from $\Lambda_{\rm NP}$ to
$\Lambda_{\rm EW}$ in the MSSM. The input and output values of
other relevant parameters can be found from Table II.}
\end{center}
\end{figure}


\newpage

\begin{thebibliography}{99}
\bibitem{SNO} SNO Collaboration, Q.R. Ahmad {\it et al.},
Phys. Rev. Lett. {\bf 89}, 011301 (2002).

\bibitem{SK} For a review, see: C.K. Jung {\it et al.},
Ann. Rev. Nucl. Part. Sci. {\bf 51}, 451 (2001).

\bibitem{KM} KamLAND Collaboration, K. Eguchi {\it et al.},
Phys. Rev. Lett. {\bf 90}, 021802 (2003); CHOOZ Collaboration, M.
Apollonio {\it et al.}, Phys. Lett. B {\bf 420}, 397 (1998); Palo
Verde Collaboration, F. Boehm {\it et al.}, Phys. Rev. Lett. {\bf
84}, 3764 (2000).

\bibitem{K2K} K2K Collaboration, M.H. Ahn {\it et al.},
Phys. Rev. Lett. {\bf 90}, 041801 (2003).

\bibitem{PDG} Particle Data Group, S. Eidelman {\it et al.}, Phys.
Lett. B {\bf 592}, 1 (2004).

\bibitem{Vissani} A. Strumia and F. Vissani, hep-ph/0503246.

\bibitem{Review} For recent reviews with extensive references,
see: H. Fritzsch and Z.Z. Xing, Prog. Part. Nucl. Phys. {\bf 45},
1 (2000); S.F. King, Rept. Prog. Phys. {\bf 67}, 107 (2004); G.
Altarelli and F. Feruglio, New J. Phys. {\bf 6}, 106 (2004).

\bibitem{HPS} P.F. Harrison, D.H. Perkins, and W.G. Scott, Phys.
Lett. B {\bf 530}, 167 (2002); Z.Z. Xing, Phys. Lett. B {\bf 533},
85 (2002); P.F. Harrison and W.G. Scott, Phys. Lett. B {\bf 535},
163 (2002); X.G. He and A. Zee, Phys. Lett. B {\bf 560}, 87
(2003); E. Ma, Phys. Rev. Lett. {\bf 90}, 221802 (2003); C.I. Low
and R.R. Volkas, Phys. Rev. D {\bf 68}, 033007 (2003); N. Li and
B.Q. Ma, Phys. Rev. D {\bf 71}, 017302 (2005); G. Altarelli and F.
Feruglio, Nucl. Phys. B {\bf 720}, 64 (2005); F. Plentinger and W.
Rodejohann, hep-ph/0507143.

\bibitem{Ma} E. Ma, Phys. Rev. D {\bf 70}, 031901 (2004);
New J. Phys. {\bf 6}, 104 (2004); hep-ph/0508099.

\bibitem{A4} G. Altarelli and
F. Feruglio, in Ref. \cite{HPS}; and references therein.

\bibitem{He} K.S. Babu and X.G. He, hep-ph/0507217.

\bibitem{Zee} A. Zee, hep-ph/0508278.

\bibitem{Grimus} See, e.g., W. Grimus, A.S. Joshipura,
S. Kaneko, L. Lavoura, and M. Tanimoto, JHEP {\bf 0407}, 078
(2004); W. Grimus and L. Lavoura, JHEP {\bf 0508}, 013 (2005); J.
Phys. G {\bf 31}, 693 (2005); Nucl. Phys. B {\bf 713}, 151 (2005);
M. Lindner, M. Ratz, and M.A. Schmidt, hep-ph/0506280.

\bibitem{RGE} P.H. Chankowski and Z. Pluciennik, Phys. Lett. B
{\bf 316}, 312 (1993); K.S. Babu, C.N. Leung, and J. Pantaleone,
Phys. Lett. B {\bf 319}, 191 (1993); M. Tanimoto, Phys. Lett. B
{\bf 360}, 41 (1995); S. Antusch, M. Drees, J. Kersten, M.
Lindner, and M. Ratz, Phys. Lett. B {\bf 519}, 238 (2001). Only
the last reference yields the correct beta-function for the SM.

\bibitem{SS} P. Minkowski, Phys. Lett. B {\bf 67}, 421
(1977); T. Yanagida, in {\it Proceedings of the Workshop on
Unified Theory and the Baryon Number of the Universe}, edited by
O. Sawada and A. Sugamoto (KEK, Tsukuba, 1979), p. 95; M.
Gell-Mann, P. Ramond, and R. Slansky, in {\it Supergravity},
edited by F. van Nieuwenhuizen and D. Freedman (North Holland,
Amsterdam, 1979), p. 315; S.L. Glashow, in {\it Quarks and
Leptons}, edited by M. L$\rm\acute{e}vy$ {\it et al.} (Plenum, New
York, 1980), p. 707; R.N. Mohapatra and G. Senjanovic, Phys. Rev.
Lett. {\bf 44}, 912 (1980).

\bibitem{RGE1} For recent comprehensive works,
see: J.A. Casas, J.R. Espinosa, A. Ibarra, and I. Navarro, Nucl.
Phys. B {\bf 573}, 652 (2000); P.H. Chankowski and S. Pokorski,
Int. J. Mod. Phys. A {\bf 17}, 575 (2002); S. Antusch, J. Kersten,
M. Lindner, M. Ratz, M.A. Schmidt, JHEP {\bf 0503}, 024 (2005);
J.W. Mei, Phys. Rev. D {\bf 71}, 073012 (2005).

\bibitem{Threshold} See, e.g., J.A. Casas, J.R. Espinosa, A.
Ibarra, and I. Navarro, Nucl. Phys. B {\bf 556}, 3 (1999); Nucl.
Phys. B {\bf 569}, 82 (2000); S. Antusch, M. Drees, J. Kersten, M.
Lindner, and M. Ratz, Phys. Lett. B {\bf 525}, 130 (2002); S.
Antusch, J. Kersten, M. Lindner, and M. Ratz, Phys. Lett. B {\bf
538}, 87 (2002); J.W. Mei and Z.Z. Xing, Phys. Rev. D {\bf 70},
053002 (2004); S. Antusch, J. Kersten, M. Lindner, M. Ratz, M.A.
Schmidt, in Ref. \cite{RGE1}; J.W. Mei, in Ref. \cite{RGE1}; J.
Ellis, A. Hektor, M. Kadastik, K. Kannike, and M. Raidal,
hep-ph/0506122.

\bibitem{Antusch} S. Antusch, J. Kersten, M. Lindner, and M. Ratz,
Nucl. Phys. B {\bf 674}, 401 (2003).

\bibitem{Luo} S. Luo, J. Mei, and Z.Z. Xing, hep-ph/0507065; Phys.
Rev. D (in press).

\bibitem{FX96} H. Fritzsch and Z.Z. Xing, Phys. Lett. B {\bf 372},
265 (1996); Phys. Lett. B {\bf 440}, 313 (1998); Phys. Rev. D {\bf
61}, 073016 (2000); Phys. Lett. B {\bf 598}, 237 (2004).

\bibitem{Haba} N. Haba, Y. Matsui, N. Okamura, and M. Sugiura,
Prog. Theor. Phys. {\bf 103}, 145 (2000); N. Haba, N. Okamura, and
M. Sugiura, Prog. Theor. Phys. {\bf 103}, 367 (2000); N. Haba, Y.
Matsui, and N. Okamura, Eur. Phys. J. C {\bf 17}, 513 (2000).

\bibitem{J} C. Jarlskog, Phys. Rev. Lett. {\bf 55}, 1039 (1985).

\bibitem{Wang} See, e.g., Y.F. Wang, hep-ex/0411028.

\bibitem{FX01} H. Fritzsch and Z.Z. Xing, Phys. Lett. B {\bf 517},
363 (2001); Z.Z. Xing, Int. J. Mod. Phys. A {\bf 19}, 1 (2004).

\bibitem{Xing03} Z.Z. Xing, J. Phys. G {\bf 29}, 2227 (2003); C.
Giunti, hep-ph/0209103 (unpublished).
\end{thebibliography}
\end{document}